\documentclass[aps,prd,twocolumn,showpacs,preprintnumbers,amsmath,amssymb]{revtex4}

\begin{document}


\title{The Gravitational-Electromagnetic Analogy: 
A Possible Solution to the Vacuum-Energy and Dark-Energy Problems}


\author{Richard J.\ Cook}
\email[]{richard.cook@usafa.edu}
\affiliation{Department of Physics, U.S. Air Force 
Academy, Colorado Springs, CO 80840-5701}

\date{\today}

\begin{abstract}
There is a set of first-order differential equations for the curvature tensor in general relativity (the ``curvature equations'' or CEs for short) that are strikingly similar to the Maxwell equations of electrodynamics.  This paper considers whether Mother Nature may have used the same basic pattern for her laws of gravitation and electrodynamics, in which case the CEs might be viewed as the fundamental field equations of gravitation in place of Einstein's equation.  This is \emph{not} a new theory of gravitation (because the CEs are derivable from Einstein's equation), but rather is a mild reinterpretation of general relativity that solves the vacuum-energy problem and the dark-energy problem of cosmology.  We show that: (1) any solution of Einstein's equation is also a solution of the CEs; (2) the CEs solve the vacuum-energy problem in as much as the vacuum energy of quantum fields has no effect on the curvature or the metric of spacetime, according to these equations; (3) if conditions implied by the principle of equivalence are obeyed, then the CEs are equivalent to an Einstein equation with an additional contribution to the energy-momentum tensor that takes the form of a cosmological term, but the cosmological constant, in this context, is a free parameter unrelated to the vacuum energy of quantum fields; and (4) Einstein's field equation with cosmological term turns out to be a first integral of the curvature equations, and the cosmological constant is an integration constant determined by the initial conditions for the universe.     These results allow one to understand, in a natural way, how the effective energy density of the observed cosmological constant can be so vastly smaller than  estimates of the vacuum energy of quantum fields and why the vacuum energy of quantum fields does not contribute as a source of curvature.
\end{abstract}

\pacs{04.20.-q, 04.20.Cv, 04.20.Dw, 04.90.+e, 95.35.+d, 98.80.-k, 98.80.Bp }

\maketitle

\section{Introduction\label{sec:Intro}}
There exits a close formal analogy between certain equations for the 
curvature tensor in general relativity and the equations of classical 
electrodynamics.  Using the Einstein field equations (with or without 
the cosmological term) and the Bianchi identity, it is easy to show 
that the curvature tensor ${R^{\alpha}}_{\beta \mu \nu}$ obeys the 
``curvature equations'' (or CEs for short)
\begin{subequations}
\label{CEs}
\begin{equation}
 \nabla_{\omega}R_{\alpha \beta \mu \nu}+
 \nabla_{\nu}R_{\alpha \beta \omega \mu}+
  \nabla_{\mu}R_{\alpha \beta \nu \omega}=0 ,  
	\label{Bianchi}
\end{equation}
\begin{equation}
\nabla_{\nu}{R_{\alpha \beta}}^{\mu \nu}=-4\pi{J_{\alpha 
\beta}}^{\mu},
\label{SourceEq}
\end{equation}
\end{subequations}
where (\ref{Bianchi}) is the Bianchi identity itself and the source 
term ${J_{\alpha \beta}}^{\mu}$ is determined by the energy-momentum 
tensor ${T_{\beta}}^{\mu}$ and its derivatives as
\begin{subequations}
\label{CEsource}
\begin{equation}
{J_{\alpha \beta}}^{\mu}=\frac{2G}{c^{4}}
\left[ 
\nabla_{\alpha}\bar{T}_{\beta}^{\ \mu}
-\nabla_{\beta}\bar{T}_{\alpha}^{\ \mu}
 \right],
\label{SourceTerm}
\end{equation}
where
\begin{equation}
\bar{T}_{\alpha}^{\ \mu}=T_{\alpha}^{\ \mu}-\frac{1}{2}\delta_{\alpha}^{\ \mu}T
\label{HilbertC}
\end{equation}
\end{subequations}
is the Hilbert conjugate of $T_{\alpha}^{\ \mu}$ and $T=T_{\mu}^{\ \mu}$.  (see Appendix \ref{sec:CEderivation} for a derivation).

The curvature equations (\ref{CEs}) are formally analogous to Maxwell's equations 
\begin{subequations}
\begin{equation}
\nabla_{\omega}F_{\mu \nu}+\nabla_{\nu}F_{\omega 
\mu}+\nabla_{\mu}F_{\nu \omega}= 0,
\label{Max1}
\end{equation}
\begin{equation}
\nabla_{\nu}F^{\mu \nu}=-4\pi J^{\mu}.
\label{Max2}
\end{equation}
\end{subequations}
In fact, for each independent choice of the indices $\alpha \beta$ (= 
01, 02, 03, 12, 13, and 23), equations (\ref{CEs})
are formally identical to Maxwell's equations.  The formal 
similarity between Maxwell's equations and the CEs has been noted by 
others \cite{Yang1, Szczyrba, Hehl, Dolan}.

The analogy goes further.  As emphasized by Misner, Thorne and 
Wheeler \cite{Misner}, the gravitational analogue of the Lorentz force law
\begin{equation}
\frac{D^{2}x^{\alpha}}{d\tau^{2}}=\frac{q}{m}{F^{\alpha}}_{\beta}
\frac{dx^{\beta}}{d\tau},
\label{Lorentz}
\end{equation}
is not the geodesic equation but the equation of geodesic deviation
\begin{equation}
\frac{D^{2}\xi^{\alpha}}{d\tau^{2}}={R^{\alpha}}_{\beta \nu \mu}
\frac{dx^{\beta}}{d\tau} \frac{dx^{\nu}}{d\tau} \xi^{\mu},
\label{Deviation}
\end{equation}
where $\xi^{\alpha}$ is the separation between neighboring geodesics 
and $D/d\tau$ the absolute derivative.  Here again we see the similar 
roles played by ${F^{\alpha}}_{\beta}$ and ${R^{\alpha}}_{\beta \mu 
\nu}$ in the two theories.

The analogue of the vector 
potential $A_{\mu}$, which determines the electromagnetic field tensor
\begin{equation}
F_{\mu \nu}= \partial_{\mu}A_{\nu}-\partial_{\nu}A_{\mu},
\label{FieldTensor}
\end{equation}
is the set of affine connection coefficients $\Gamma^{\alpha}_{\ \beta \mu}$, which 
determines the curvature tensor
\begin{equation}
{R^{\alpha}}_{\beta \mu \nu} =
\partial_{\mu}\Gamma^{\alpha}_{\ \beta \nu}-
\partial_{\nu}\Gamma^{\alpha}_{\ \beta \mu}+
\Gamma^{\alpha}_{\ \tau \mu}\Gamma^{\tau}_{\ \beta \nu}-
\Gamma^{\alpha}_{\ \tau \nu}\Gamma^{\tau}_{\ \beta \mu}.
\label{CurveTens}
\end{equation}
Note the similarity of the first two terms in (\ref{CurveTens}) and the terms in (\ref{FieldTensor})
[the non-linear terms in (\ref{CurveTens}) presumably being associated 
with the fact that the (pseudo) energy of the gravitational field is 
itself a source of gravitational field].

Even the geometric interpretation of the curvature tensor has an 
electromagnetic analogue.  The potentials $A_{\mu}$ 
may be viewed as connection coefficients on a principle fiber bundle, 
and the $F_{\mu \nu}$s are then the curvatures associated with these 
connections \cite{Yang,Drechsler}.

We may also note that the curvature equations are derivable from an action principle that is analogous to the one for classical electrodynamics.  In electrodynamics the field tensor
\begin{equation}
 F_{\mu\nu}=\nabla_{\mu}A_{\nu}-\nabla_{\nu}A_{\mu}
 \label{FieldTensor2}
 \end{equation}
  is derivable from a vector potential $A_{\mu}$.  This is the content of, and the solution to, the first Maxwell equation (\ref{Max1}).  Then, for arbitrary but small variations $\delta A_{\mu}$ of the vector potential, the action for the electromagnetic field \cite{Landau1},
\begin{equation}
S_{EM}=\frac{-1}{16\pi} \int_{\Sigma}  \left(F^{\mu\nu}F_{\mu\nu}+16\pi J^{\nu}A_{\nu}\right)\sqrt{-g}\ d^{4}x,
\label{EMaction}
\end{equation}
is stationary ($\delta S_{EM}=0$), if and only if $F^{\mu\nu}$ satisfies the second Maxwell equation (\ref{Max2}).  

On the basis of the gravitational-electromagnetic analogy, we expect that the action for the gravitational field $R^{\alpha}_{\ \beta\mu\nu}$, for a prescribed source term $J^{\mu}_{\ \alpha\beta}$, is the analogue of the electromagnetic action (\ref{EMaction}), namely
\begin{equation}
S_{G}=\frac{-1}{16\pi}\int_{\Sigma} \left(R^{\alpha\beta\mu\nu}R_{\alpha\beta\mu\nu}+16\pi J_{\mu}^{\ \alpha\beta}\Gamma^{\mu}_{\ \alpha\beta}\right)\sqrt{-g}\ d^{4}x.
\label{GravAction}
\end{equation}
A gravitational field $R^{\alpha}_{\ \beta\mu\nu}$ derived from ``potentials''  $\Gamma^{\mu}_{\ \alpha\beta}$ by means of Eq.~(\ref{CurveTens}) is a solution of the first curvature equation (\ref{Bianchi}), just as a field tensor $F_{\mu\nu}$ derived from a vector potential $A_{\nu}$ by means of Eq.~(\ref{FieldTensor}) is a solution of the first Maxwell equation (\ref{Max1}).  Continuing to follow the gravitational-electromagnetic analogy, we calculate the variation $\delta S_{G}$ of (\ref{GravAction}) resulting from a variation $\delta\Gamma^{\mu}_{\ \alpha\beta}$ of the ``potentials''  $\Gamma^{\mu}_{\ \alpha\beta}$, treating these as variables independent of the metric $g_{\alpha\beta}$ (a Palatini variation).  The variational principle $\delta S_{G}=0$,  for small but arbitrary   $\delta\Gamma^{\mu}_{\ \alpha\beta}$  vanishing on the boundary  $\partial\Sigma$ of the 4-volume $\Sigma$ of interest, yields the second curvature equation (\ref{SourceEq}), as demonstrated in Appendix \ref{sec:Action}.  The key relation in this derivation is the Palatini identity
\begin{equation}
\delta R^{\alpha}_{\ \beta\mu\nu}=\nabla_{\mu}\left(\delta\Gamma^{\alpha}_{\ \beta\nu}\right)-\nabla_{\nu}\left(\delta\Gamma^{\alpha}_{\ \beta\mu}\right),
\label{PalatiniIdentity}
\end{equation}
which is exactly analogous to the relation
\begin{equation}
\delta F_{\mu\nu}=\nabla_{\mu}\left(\delta A_{\nu}\right)-\nabla_{\nu}\left(\delta A_{\mu}\right)
\label{EMvariation}
\end{equation}
in electrodynamics.  We must caution the reader that this variational calculation is not entirely satisfactory.  It works only if $J_{\alpha\beta}^{\ \ \ \mu}$ is treated as a \emph{prescribed source} independent of the connection coefficients $\Gamma^{\mu}_{\ \alpha\beta}$.  We have not yet found an action principle from which the curvature equations, Maxwell's equations, and particle equations of motion can be derived in a fully consistent manner.

Not all of the results derived from the curvature equations are \emph{exactly} analogous to those derived from Maxwell's equations because the field tensors $R^{\alpha}_{\ \beta\mu\nu}$ and $F_{\mu\nu}$ are of different rank, but the similarities are nevertheless often striking.  Covariant derivatives acting on the higher rank curvature equations frequently generate more terms in a result than in the analogous calculation based on the Maxwell equations.  For example, the wave equation for $F_{\mu\nu}$ in curved spacetime is
\begin{eqnarray}
\square^{2}F_{\mu\nu}&=&  4\pi \left(\nabla_{\mu}J_{\nu}-\nabla_{\nu}J_{\mu}\right)  \nonumber\\
 &   & \mbox{ } + R^{\sigma}_{\ \mu}F_{\sigma\nu}-R^{\sigma}_{\ \nu}F_{\sigma\mu}  \nonumber \\
 &   & \mbox{ } + \left(R^{\sigma\ \tau}_{\ \mu\ \nu}-R^{\sigma\ \tau}_{\ \nu\ \mu}\right)F_{\tau\sigma} , \label{MaxWave}
\end{eqnarray}
whereas the wave equation for $R^{\alpha}_{\ \beta\mu\nu}$, derived from the curvature equations by essentially the same steps, reads
\begin{eqnarray}
\square^{2}R^{\alpha}_{\ \beta\mu\nu}&=&4\pi\left(\nabla_{\mu}J^{\alpha}_{\ \beta\nu}-\nabla_{\nu}J^{\alpha}_{\ \beta\mu}\right) \nonumber \\
& & \mbox{ } +R^{\sigma}_{\ \mu}R^{\alpha}_{\ \beta\sigma\nu}-R^{\sigma}_{\ \nu}R^{\alpha}_{\ \beta\sigma\mu} \nonumber \\
& & \mbox{ } +\left(R^{\sigma\ \tau}_{\ \mu\ \nu}-R^{\sigma\ \tau}_{\ \nu\ \mu}\right)R^{\alpha}_{\ \beta\tau\sigma} \nonumber \\
& & \mbox{ } +2\left(R^{\alpha\ \tau}_{\ \sigma\ \nu}R^{\sigma}_{\ \beta\nu\tau}-R^{\sigma\ \tau}_{\ \sigma\ \nu}R^{\sigma}_{\ \beta\nu\tau}\right) .
\label{CurvatureWave}
\end{eqnarray}
Here only the last line deviates from an exact formal analogue of Eq.~(\ref{MaxWave}).

The above results amply illustrate the striking similarity between the curvature equations and Maxwell's equations, but there is one place in \emph{conventional general relativity} (CGR) where the gravitational-electromagnetic analogy breaks down.  In electrodynamics the Maxwell equations are viewed as the fundamental equations of the theory, but in CGR, the analogue of Maxwell's equations, namely the curvature equations, are not viewed as the fundamental equations of gravitation.  Rather, the Einstein equation $G^{\mu\nu}=8\pi GT^{\mu\nu}/c^{4}$ (either with or without the cosmological term) plays this role.

This raises the intriguing question ``Is it possible that Mother Nature has chosen the same basic pattern for her equations of gravitation and electrodynamics?''  That is to say, is it reasonable to view the curvature equations (\ref{CEs}) as the field equations of gravitation in place of Einstein's equation?  By postulating that the CEs are the proper field equations of general relativity, we complete the formal gravitational-electromagnetic analogy and we find that the vacuum energy problem and the dark energy problem have natural solutions within this framework.

It must be emphasized at the outset that this is \emph{not} a new theory of gravitation.  As shown in Appendix A, the curvature equations are derivable from the conventional Einstein equation, with or without the cosmological term, and consequently are true relations in conventional general relativity.  The present paper does nothing more than suggest that, among the correct equations of conventional general relativity, the curvature equations might better serve as the field equations of the theory than the Einstein equation.  Furthermore, the choice of the curvature equations as the field equations of general relativity does not abandon the Einstein equation, which is recovered as the first integral of the curvature equations, but with a new interpretation for the cosmological constant and no contribution from the vacuum energy of quantum fields.  These changes solve the vacuum-energy problem and the dark-energy problem while keeping the Einstein equation as the \emph{working} field equations of general relativity..   For these reasons, the present work is properly viewed as a mild reinterpretation of general relativity, and not as an alternative theory of gravitation.

In our work with the curvature equations, we adhere to the principle of equivalence and the principle of general covariance (the fundamental postulates of general relativity).  In view of the various proofs that the Einstein equation follows from these postulates, one might wonder how we can seriously consider a different set of field equations for general relativity.  The answer is that Einstein's equation does not follow from these postulates alone, but from these postulates plus the lower status assumption that the field equations should contain no higher than second-order derivatives of the metric tensor (the curvature equations contain third-order derivatives of the metric, and thus are ruled out by this assumption).  Here we drop the ``second-order assumption'', and instead follow the gravitational-electromagnetic analogy in our selection of field equations.  This does not mean that we will continually be dealing with third-order partial differential equations (which, admittedly, is distasteful to many physicists) because it turns out that the second-order Einstein equation, with cosmological term determined by initial conditions and no contribution from vacuum energy, is a first integral of the curvature equations, and it is this equation that one routinely works with to obtain solutions of the curvature equations. 

In the following section we define, for the purpose of this paper, what we mean by the terms ``vacuum-energy problem'', ``dark-energy problem'', and ``cosmological-constant problem''.  In Section \ref{sec:Properties} we discuss various properties of the CEs, and show that: (1) any solution of the conventional Einstein equation is also a solution of the CEs; (2) according to the CEs, the vacuum energy of quantum fields is not a source of curvature, and has no effect on the metric of spacetime; and (3) the CEs have the correct Newtonian limit.
Section \ref{sec:FirstInteg} shows that the first integral of the curvature equations is an Einstein equation in which the energy-momentum tensor is augmented by an ``integration tensor'' (a set of integration functions), which obey certain condition.  The dark-energy problem is addressed in Section \ref{sec:Solution}, where it is shown that the integration tensor for the CEs takes the form of a cosmological term, when reasonable conditions are imposed on the vacuum state, and the cosmological constant in this Einstein equation does not represent the energy density of the quantum vacuum.     The paper concludes in Section \ref{sec:Conclusion} with comments on the central results of the paper and on the shortcomings of this answer to the dark-energy and vacuum-energy puzzles. 

\section{\label{sec:DarkEnergy}The Vacuum-energy, Dark-Energy, and Cosmological-Constant Problems}

There is some confusion in the literature as to what exactly is meant by the terms ``vacuum-energy problem'', ``dark-energy problem'', and ``cosmological-constant problem.'' In this section we define what \emph{we} mean by these terms for the purpose of the present paper.

The \emph{vacuum-energy problem} has a long history.  As soon as it became evident that quantization of the electromagnetic field implies a large (perhaps infinite) energy density for the vacuum, it was clear that this poses a problem for general relativity; for if reasonable estimates of the energy density $\rho_{qv}c^{2}$ of the quantum vacuum (qv) are taken at face value and used as the source term in Einstein's equation, the resulting spacetime metric is violently at variance with experience \cite{Feynman}.  But there were many divergent quantities in the young quantum electrodynamics (QED), and so the gravitational problem was largely ignored until: (1) the advent of renormalization theory, which showed that QED is a viable theory, and (2) the prediction by Casimir and the subsequent experimental demonstration that the vacuum energy has measurable consequences \cite{Casimir, Sparnaay}, indicating that the vacuum energy cannot simply be ignored.  However, renormalization theory did not solve the vacuum-energy problem for gravitation.  Today estimates of the vacuum energy-momentum tensor of quantum fields still give nonsensical results when used as source term in Einstein's equation, and the current practice in general relativity is usually to simply ignore this immense contribution to the energy-momentum tensor when solving Einstein's equation
~\cite{NormalOrder}. For the purpose of this paper we define the \emph{vacuum-energy problem} narrowly as the following question.

\begin{quote}
\textbf{Vacuum-Energy Problem:}

 \textit{Why is it that the immense vacuum energy density of quantum fields does not contribute to the energy-momentum tensor as a source of curvature in Einstein's equation?}
\end{quote}

The phrase \emph{cosmological-constant problem} has an even longer history than the vacuum energy problem.  In 1917 the cosmological term $\Lambda g^{\mu\nu}$ was  added by Einstein to his field equation as
\begin{equation}
G^{\mu\nu}+\Lambda g^{\mu\nu}=\frac{8\pi G}{c^{4}} T^{\mu\nu},
\label{EinsteinEq}
\end{equation}
in order permit a solution representing a static universe \cite{Einstein1}, but was famously rejected by him as his ``greatest blunder'' when he became aware of Hubble's discovery of the expanding universe [Eddington's later demonstration that the cosmological term does not give \emph{stable} static solutions of Eq.~(\ref{EinsteinEq}) also argues against a static universe of the Einstein type \cite{Eddington}].  After this, many cosmologists were content to invoke Ockham's razor and work with an Einstein equation not containing a cosmological term (a notable exception to this thinking is found in the work of Carroll, Press, and Turner who anticipated the need for a cosmological term\cite{Carroll} when it was not popular).  We define the phrase \emph{cosmological-constant problem} to be the questions:

\begin{quote}
\textbf{Cosmological-Constant Problem:}

\textit{Should there be or should there not be a cosmological term in Einstein's equation?  And, if there should be such a term in Einstein's equation, what is its physical interpretation and how is its value to be predicted?}
\end{quote}

There was an abrupt change in the conventional wisdom concerning the cosmological term in recent years resulting from the high-redshift supernovae observations of Perlmutter et. al. \cite{Perlmutter}, Riess et.al. \cite{Riess}, Tonry et.al.~\cite{Tonry}, and Knop et.al.~\cite{Knop}; and corroborated by the cosmic microwave background observations of Spergel et.al.~\cite{Spergel} and Sievers et.al.~\cite{Sievers},  showing that the universe is in a state of accelerated expansion. The accelerating expansion can be ``explained'' by a cosmological term in Einstein's equation with a certain value for the cosmological constant, but the physical origin if such a term is not understood.

What seems clear at the present time is that agreement with cosmological observations is achieved when we use an Einstein equation of the form
\begin{equation}
G^{\mu\nu}=\frac{8\pi G}{c^{4}}\left(T^{\mu\nu}_{c}+T^{\mu\nu}_{d}\right),
\label{DarkDensityEq}
\end{equation}
in which: (1) the immense vacuum energy-momentum of quantum fields $T^{\mu\nu}_{qv}=-\rho_{qv}c^{2}g^{\mu\nu}$ is ignored; (2) the ``classical'' energy-momentum tensor $T^{\mu\nu}_{c}$ of known matter and radiation, which at the present time consists of a pressureless ``dust'' of galaxies and the cosmic microwave background radiation (CMBR), is included; and (3) a ``dark-energy'' energy-momentum tensor of the form
\begin{equation}
T^{\mu\nu}_{d}=-\rho_{d}c^{2}g^{\mu\nu}, 
\label{DarkEMTensor}
\end{equation}
with effective mass density $\rho_{d}=\Omega_{d}\rho_{cr}$ with $\Omega_{d}=0.7\pm 0.1$, is added to the source term of Einstein's equation to account for the accelerated expansion of the universe.  Here $\rho_{cr}$ is the critical density,
\begin{equation}
\rho_{cr}=\frac{3H^{2}_{o}}{8\pi G}=1.88\times 10^{-29}\ h^{2}\ \frac{g}{cm^{2}},
\label{Critical}
\end{equation}
where $h\equiv H_{0}/[100(km/s)/Mpc]$ and $H_{0}$ is the Hubble constant [we are using a metric with signature (-1, 1, 1, 1)].  From the point of view of quantum field theory, the ``classical'' energy-momentum tensor $T^{\mu\nu}_{c}$ represents excitations of  quantum fields above the vacuum state, i.e., material particles and field excitations such as photons.

The ``dark-energy'' tensor $T^{\mu\nu}_{d}$ has the same form as the energy-momentum tensor of the quantum vacuum $T^{\mu\nu}_{qv}=-\rho_{qv}c^{2}g^{\mu\nu}$, which is also the same as the equivalent energy-momentum tensor $T^{\mu\nu}_{\Lambda}$ of a cosmological term,
\begin{equation}
T^{\mu\nu}_{\Lambda}=-\frac{\Lambda c^{4}}{8\pi G}g^{\mu\nu}=-\rho_{\Lambda}c^{2}g^{\mu\nu}.
\label{LambdaVacEnergy}
\end{equation}
This has lead to the popular surmise that the dark energy and the cosmological term are one and the same, $\rho_{d}=\rho_{\Lambda}=\Lambda c^{2}/8\pi G$, and that $T^{\mu\nu}_{d}$ is perhaps the vacuum energy-momentum tensor of some quantum field (or combination of quantum fields) not yet understood.  There are many proposed models for the dark energy that differ from the locally Lorentz invariant form $-\rho c^{2}g^{\mu\nu}$ (with $\rho$ a constant) ranging from scalar field models, to higher-dimensional models, to various modifications of general relativity \cite{Peebles}.  We do not consider these models in this paper because the CEs lead naturally to a cosmological or vacuum-energy like term (\ref{DarkEMTensor}) in Einstein's equation, and this is sufficient for our present purpose.

The surprise in fitting Eq.~(\ref{DarkDensityEq}) to observation is that the dark-energy density $\rho_{d}c^{2}$ turns out to be vastly smaller than estimates of the vacuum energy density of quantum fields; $\rho_{d}c^{2}$ being smaller than these estimates by as much as 120 orders of magnitude.  For our purpose, the dark-energy problem is the question:

\begin{quote}
\textbf{Dark-Energy Problem:} 

\textit{What is the dark energy, and how can its density $\rho_{d}c^{2}$ be so vastly smaller than estimates of the vacuum energy density of quantum fields?}
\end{quote}

For the purpose of the following discussion, we define \emph{conventional general relativity} (or CGR for short) as Einstein's theory with the \emph{full} energy-momentum tensor $T^{\mu\nu}$ acting as the source term in Einstein's equation,
\begin{equation}
G^{\mu\nu}=\frac{8\pi G}{c^{4}}T^{\mu\nu}.
\label{StandardEquation}
\end{equation} 
We take the full energy-momentum tensor $T^{\mu\nu}$, including the vacuum energy of quantum fields, as the proper source term for the conventional Einstein equation because there does not appear to be any logically satisfactory way to keep only certain desirable energy-momentum tensors to the exclusion of others we might prefer to ignore. 
The \emph{full} energy-momentum tensor is understood to be the classical energy-momentum tensor $T^{\mu\nu}_{c}$ together with the energy-momentum tensor of the quantum vacuum $T^{\mu\nu}_{qv}$ and the effective energy-momentum tensor $T^{\mu\nu}_{\Lambda}$ associated with a cosmological term, if the latter represents an energy-momentum tensor different from $T^{\mu\nu}_{qv}$,
\begin{equation}
T^{\mu\nu}=T^{\mu\nu}_{c}+T^{\mu\nu}_{qv}+T^{\mu\nu}_{\Lambda}.
\label{TotalTensor}
\end{equation}
We do not explicitly include a dark-energy term $T^{\mu\nu}_{d}=-\rho_{d}c^{2}g^{\mu\nu}$ here because, we do not believe this is the energy-momentum tensor of any field, but rather has an entirely different interpretation based on the curvature equations.

In the following sections, we show that the curvature equations answer the vacuum-energy problem, the cosmological-constant problem, and the dark-energy problem in a natural way, and provide insight into the nature of dark energy.

\section{\label{sec:Properties} PROPERTIES OF THE CURVATURE EQUATIONS}

Let us formally introduce our postulate for the reinterpretation of general relativity.

\begin{quote}
\textbf{Postulate:}  \textit{The spacetime metric} $g_{\alpha\beta}$ \textit{obeys the curvature field equations,}
\begin{subequations}
\label{CE2s}
\begin{equation}
 \nabla_{\omega}R_{\alpha \beta \mu \nu}+
 \nabla_{\nu}R_{\alpha \beta \omega \mu}+
  \nabla_{\mu}R_{\alpha \beta \nu \omega}=0 ,  
	\label{Bianchi2}
\end{equation}
\begin{equation}
\nabla_{\nu}{R_{\alpha \beta}}^{\mu \nu}=-4\pi{J_{\alpha 
\beta}}^{\mu},
\label{SourceEq2}
\end{equation}
\end{subequations}
\textit{where the source term} $J_{\alpha\beta}^{\ \ \mu}$ \textit{is determined by the full energy-momentum tensor} $T^{\mu\nu}=T^{\mu\nu}_{c}+T^{\mu\nu}_{qv}+T^{\mu\nu}_{\Lambda}$ \textit{and its derivatives as}
\begin{equation}
{J_{\alpha \beta}}^{\mu}=\frac{2G}{c^{4}}
\left(
\nabla_{\alpha}\bar{T}_{\beta}^{\mu}
-\nabla_{\beta}\bar{T}_{\alpha}^{\mu} \right),
\label{SourceTerm2}
\end{equation}
\textit{where $\bar{T}_{\alpha}^{\ \mu}=T_{\alpha}^{\ \mu}-\delta_{\alpha}^{\ \mu}T/2$ is the Hilbert conjugate of the energy-momentum tensor $T_{\alpha}^{\ \mu}$.  The curvature equations are viewed as the fundamental field equations of general relativity, replacing the conventional Einstein equation for the purpose of predicting the geometry of spacetime.}
\end{quote}

These equations determine the curvature tensor $R^{\alpha}_{\ \beta\mu\nu}$ generated by a given source term $J_{\alpha\beta}^{\ \ \mu}$ in a manner analogous to the way Maxwell's equations determine the electromagnetic field tensor $F_{\mu\nu}$ generated by a given current density $J^{\mu}$.  But the CEs do not determine the metric tensor $g_{\alpha\beta}$ uniquely.   Because these equations are generally covariant, the metric $g_{\bar{\alpha}\bar{\beta}}$ obtained from $g_{\alpha\beta}$ by a coordinate transformation is also a solution of the curvature equations, and so, just as for the conventional Einstein equation, there is a fourfold ambiguity in the solutions of the CEs corresponding to the four arbitrary functions  $x^{\bar{\mu}}=x^{\bar{\mu}}(x^{\nu})$ of a coordinate transformation.

Again following the gravitational-electromagnetic analogy, we need a particle equation of motion analogous to the Lorentz force law to complete the theory.  We could surmise from the principle of equivalence that the geodesic equation is this equation of motion, but it is more convenient for our present purpose to apply the principle of equivalence to the energy-momentum tensor.  In a local inertial reference frame, the  conservation of energy and momentum is described by the law $\partial_{\nu}T^{\mu\nu}=0$.  The principle of equivalence then suggests that the corresponding generally covariant law in curved spacetime is obtained by replacing the partial derivative $\partial_{\nu}$ by the covariant derivative $\nabla_{\nu}$ (the so-called comma-to-semicolon rule in a different notation). Thus we expect that $\nabla_{\nu}T^{\mu\nu}=0$ is the correct expression for local energy and momentum conservation in curved spacetime, as it is in CGR.  But the $\partial_{\nu}\rightarrow\nabla_{\nu}$ rule is not \emph{always} valid, as, for example, when transforming the wave equation for the vector potential in electrodynamics \cite{Misner2}.  Therefore we elevate this plausible surmise to a condition that must be satisfied in order for our subsequent conclusions to apply:

\begin{quote}
\textbf{Condition 1:} \textit{The energy-momentum tensor} $T^{\mu\nu}$ \textit{for matter, radiation, and/or vacuum obeys the local conservation law}
\begin{equation}
\nabla_{\nu}T^{\mu\nu}=0
\label{Conservation}
\end{equation}
\textit{for energy and momentum.}
\end{quote}

As is well known from conventional general relativity, this condition contains the geodesic equation
\begin{equation}  
\frac{d^{2}x^{\mu}}{d\tau^{2}}+\Gamma^{\mu}_{\ \alpha\beta}\frac{dx^{\alpha}}{d\tau}\frac{dx^{\beta}}{d\tau}=0
\label{Geodesic}
\end{equation}
as a special case derived by using the energy-momentum tensor for a single particle in Eq.~(\ref{Conservation}).  This calculation appears as the second step in the ``problem of motion'' in conventional general relativity, where the geodesic equation is derived from the Einstein field equation, for the purpose of showing that the geodesic equation is a consequence of the field equation alone and  need not be postulated separately \cite{Einstein2, Einstein3, Infeld}.  The first step in this calculation is to show that Eq.~(\ref{Conservation}) follows from the Einstein equation, which of course it does because the Einstein equation was fashioned to be consistent with Eq.~(\ref{Conservation}) [we recall that Einstein's first attempt at a field equation for general relativity was of the form $R^{\mu\nu}=\kappa T^{\mu\nu}$ with $\kappa$ a constant, but he rejected this equation when he realized that it was not consistent with Eq.~(\ref{Conservation})].  In short, Einstein felt so strongly that Eq.~(\ref{Conservation}) must be correct that he effectively used this as a postulate in constructing his theory.  But, having derived his field equation in this way, Eq.~(\ref{Conservation}) was no longer needed as a postulate because it could then be derived from the field equation.  In the present approach, Eq.~(\ref{Conservation}) does not follow directly from the curvature equations and therefore it must be postulated separately, although its truth is extremely plausible from the principle of equivalence alone, as Einstein knew very well.

Our second condition also follows from the principle of equivalence and is familiar from conventional general relativity.  The source term $S^{\mu\nu}$ in a generic Einstein equation $G^{\mu\nu}=8\pi GS^{\mu\nu}/c^{4}$ generally consists of a part $T^{\mu\nu}_{c}$ representing all excitations of quantum fields \emph{above the vacuum state}, i.e., material particles and field excitations such as photons, and a part $S^{\mu\nu}_{v}$ describing how the vacuum acts as a source of curvature:
\begin{equation}
S^{\mu\nu}=T^{\mu\nu}_{c}+S^{\mu\nu}_{v}.
\label{FullSource}
\end{equation}
It is usually assumed that $S^{\mu\nu}_{v}$ is of the locally Lorentz invariant form
\begin{equation}
S^{\mu\nu}_{v}=-\rho_{v}c^{2}g^{\mu\nu},
\label{LEform}
\end{equation}
with $\rho_{v}$ a constant.  This form implies that the ``vacuum energy density'' $\rho_{v}c^{2}$ is the same in all local inertial reference frames, and that the ``vacuum momentum density'' $\pi^{i}=-\rho_{v}c\eta^{0i}$ vanishes in all local inertial frames.  Any other form for the vacuum source leads to different vacuum properties in different local inertial frames (different vacuum energy densities and nonzero effective vacuum momentum density in different local inertial frames), and such is clearly at variance with the principle of equivalence.  We shall refer to $S^{\mu\nu}_{v}$, namely whatever is left on the right hand side of Einstein's equation in the absence of any excitations of quantum fields (when $T^{\mu\nu}_{c}=0$), as the \emph{Einstein vacuum}, and we introduce a condition requiring the Einstein vacuum to be of the locally Lorentz invariant form, so as to be fully consistent with the principle of equivalence.

\begin{quote}
\textbf{Condition 2:}  \textit{ If the spacetime metric $g_{\mu\nu}$ obeys an Einstein equation}
\begin{equation}
G^{\mu\nu}=\frac{8\pi G}{c^{4}}\left(T^{\mu\nu}_{c}+S^{\mu\nu}_{v}\right)
\label{EinsteinWVacuum}
\end{equation}
\textit{in which $T^{\mu\nu}_{c}$ is the energy-momentum tensor of all quantum excitations above the vacuum state, then the Einstein vacuum $S^{\mu\nu}_{v}$ must be of the locally Lorentz invariant form}
\begin{equation}
S^{\mu\nu}_{v}=-\rho_{v}c^{2}g^{\mu\nu},
\label{LEform2}
\end{equation}
\textit{with $\rho_{d}$ a constant, whether or not $S^{\mu\nu}_{v}$ is the vacuum energy-momentum tensor of a quantum field.}
\end{quote}

We proceed now to study various properties of the curvature equations.  First off, we show that, according to the curvature equations, the vacuum energy-momentum tensor of a quantum field, $T^{\mu\nu}_{qv}=-\rho_{qv}c^{2}g^{\mu\nu}$, contributes nothing to the source term $J_{\alpha\beta}^{\ \ \mu}$ of the curvature equations.  For this form of energy-momentum tensor, the Hilbert conjugate $\bar{T}_{\alpha}^{\ \mu}$ of $T_{qv\ \alpha}^{\ \ \ \ \mu}$ is a constant multiple of the metric tensor $\delta_{\alpha}^{\ \mu}$, 
\begin{equation}
\bar{T}_{\alpha}^{\ \mu}=T_{qv\ \alpha}^{\ \ \ \ \mu}-\frac{1}{2}\delta_{\alpha}^{\ \mu}T_{qv}=\rho_{qv}c^{2}\delta_{\alpha}^{\ \mu},
\label{Hilbert}
\end{equation}
and, therefore, the covariant derivatives in Eq.~(\ref{SourceTerm2}) vanish, and the quantum vacuum has no effect on the solutions $g_{\mu\nu}$ of the curvature equations.  The same can be said for the effective energy-momentum tensor  $T^{\mu\nu}_{\Lambda}=-\rho_{\Lambda}c^{2}g^{\mu\nu}$ associated with the cosmological term because this is also a constant multiple of the metric.  Hence we have:

\begin{quote}
\textbf{Result I:} \textit{The vacuum energy-momentum tensor of quantum fields and the effective energy-momentum tensor of the cosmological term contribute nothing to the source term $J_{\alpha\beta}^{\ \ \mu}$ of the curvature equations, and therefore have no affect whatever on the curvature or metric of spacetime.}
\end{quote}

This clearly solves the vacuum-energy problem in as much as the immense vacuum energy density of quantum fields no longer has a catastrophic effect on the metric.  But this solution of the vacuum-energy problem raises another question: \textit{If an energy-momentum tensor of the form $T^{\mu\nu}=\kappa g^{\mu\nu}$, with $\kappa$ a constant,  does not influence the metric at all, how can the dark energy, with energy-momentum tensor of this form, cause the accelerated expansion of the universe?}  And the same question can be raised with regard to the energy-momentum tensor $T^{\mu\nu}_{fv}=-\rho_{fv}c^{2}g^{\mu\nu}$ of the ``false vacuum"  that drives inflation. The curvature equations have answers to these questions also, as shown in following sections.

Consider next whether the curvature equations are consistent with known solutions of Einstein's equation.  As shown in Appendix \ref{sec:CEderivation}, the curvature equations may be derived from the Einstein equation, with or without a cosmological term.  Therefore, any solution $g_{\alpha\beta}$ of the Einstein equation, say one that was the subject of an observational test, is also a solution of the curvature equations.

\begin{quote}
\textbf{Result II:} \textit{The curvature equations are consistent with the conventional Einstein equation in the sense that any solution of the conventional Einstein equation is also a solution of the curvature equations}
\end{quote}

This is not to say that all solutions of the curvature equations are solutions of Einstein's equation.  This is decidedly \emph{not} the case, as we shall see in the next section.  Result II merely states that there exists a solution of the curvature equations which is the same as the solution of the conventional Einstein equation for any specified energy-momentum tensor, not that this solution of the curvature equations is the one realized in nature. In fact, the solutions of the vacuum-energy and dark-energy problems provided by the curvature equations hinge on the curvature equations having solutions different from those of the conventional Einstein equation.

We wish to consider next the Newtonian limit of the curvature equations, but first we review the Newtonian limit of CGR to set the notation.  From the geodesic equation (\ref{Geodesic}) in the low-velocity, weak-field limit [where $x^{\mu}=(ct, x^{i})$ are Minkowski coordinates, $d\tau\approx dt$, and $g_{\alpha\beta}\approx\eta_{\alpha\beta}$], we identify the Newtonian acceleration due to gravity
\begin{equation}
g^{i}\equiv\frac{d^{2}x^{i}}{dt^{2}}=-c^{2}\Gamma^{i}_{\ 00}.
\label{NewtAccel}
\end{equation}
The $00$ component of the Einstein equation $R_{\mu\nu}=8\pi G(T_{\mu\nu}-g_{\mu\nu}T/2)/c^{4}$, for a time-independent field, reads $\partial_{i}\Gamma^{i}_{\ 00}=4\pi G(2T_{00}+T)/c^{4}$.  In terms of the gravitational acceleration (\ref{NewtAccel}), this is the Newtonian field equation
\begin{equation}
\nabla\cdot\mathbf{g}=-4\pi G\rho_{g},
\label{Divergence}
\end{equation}
where
\begin{equation}
\rho_{g}=\frac{2T_{00}+T}{c^{2}}
\label{ActiveGravMass}
\end{equation}
is the density of active gravitational mass \cite{Tolman} [for pressureless matter at rest,  in which case $T_{\mu\nu}=diag(\rho c^{2}, 0, 0, 0)$, $\rho_{g}$ is just the inertial mass density $\rho$].  For a time-independent field, the Christoffel symbol in (\ref{NewtAccel}) is $\Gamma^{i}_{\ 00}=-\delta^{ij}\partial_{j}(g_{00}/2)$, indicating that the gravitational acceleration is derivable from a potential $\Phi$:
\begin{eqnarray}
\mathbf{g}&=&-\nabla\Phi, \label{GradiantPhi} \\
\Phi&=&-\frac{c^{2}}{2}g_{00}. \label{Potential}
\end{eqnarray}
Using (\ref{GradiantPhi}) in Eq.~(\ref{Divergence}), we obtain the Poisson equation
\begin{equation}
\nabla^{2}\Phi=4\pi G\rho_{g}
\label{Poisson}
\end{equation}
with solution 
\begin{equation}
\Phi(\mathbf{x})=-G\int\frac{\rho_{g}(\mathbf{x}^{\prime})}{\lvert \mathbf{x}-\mathbf{x}^{\prime}\lvert }d^{3}x^{\prime}.
\label{Solution}
\end{equation}
if $\Phi(\mathbf{x})$ vanishes at spatial infinity.
Finally, the acceleration due to gravity, Eq.~(\ref{GradiantPhi}), is the negative gradient of (\ref{Solution}),
\begin{equation}
\mathbf{g}(\mathbf{x})=G\int \frac{(\mathbf{x}^{\prime}-\mathbf{x})\rho_{g}(\mathbf{x}^{\prime})}{{\lvert \mathbf{x}-\mathbf{x}^{\prime}\lvert}^{3}}d^{3}x^{\prime}.
\label{gSolution}
\end{equation}

Now for the Newtonian limit of the curvature equations, which provides insight into why the vacuum-energy density has no affect on the gravitational field.  Equation (\ref{NewtAccel}) holds true for our reinterpreted general relativity, as it does for CGR, because the geodesic equation is the same for both.  We write the curvature equation (\ref{SourceEq2}) in the form $\nabla_{\nu}R^{\alpha\ \ \nu}_{\ \beta\mu}=-4\pi J^{\alpha}_{\ \beta\mu}$.  Then, in the weak-field and static limits, the $\alpha=i$, $\beta=0$, $\mu=0$ component of this equation reads $\nabla^{2}\Gamma^{i}_{\ 00}=4\pi G\delta^{ij}\partial_{j}\rho_{g}/c^{2}$.  Hence, according to the curvature equations, the field equation for the gravitational acceleration, in Newtonian approximation, is the vector Poisson equation
\begin{equation}
\nabla^{2}\mathbf{g}=-4\pi G\nabla\rho_{g}.
\label{VectorPoisson}
\end{equation}
This equation shows that a uniform mass density ($\rho_{g}=constant$) contributes nothing to the gravitational field $\mathbf{g}$, just as the homogeneous vacuum energy $T^{\mu\nu}_{qv}=-\rho_{qv}c^{2}g^{\mu\nu}$ contributes nothing to the curvature $R^{\alpha}_{\ \beta\mu\nu}$, according to the curvature equations.

The solution of Eq.~(\ref{VectorPoisson}) that tends to zero at infinity is
\begin{equation}
 \mathbf{g}(\mathbf{x})=G\int\frac{\nabla^{\prime}\rho_{g}(\mathbf{x}^{\prime})}
 {|\mathbf{x}-\mathbf{x}^{\prime}|}d^{3}x^{\prime}.
 \label{gsolution}
 \end{equation}
 If the source density $\rho_{g}(\mathbf{x})=\rho_{c}(\mathbf{x})+\rho_{qv}$ consists of a ``classical'' mass density $\rho_{c}(\mathbf{x})$ of limited spatial extent [$\rho_{c}(\mathbf{x})\rightarrow 0$ as $|\mathbf{x}|\rightarrow\infty$], and a ``vacuum'' density $\rho_{qv}=constant$, then clearly only $\rho_{c}(\mathbf{x})$ contributes to the integral (\ref{gsolution}), and an integration by parts (with the volume integral of a gradient going to zero because $\rho_{c}(\mathbf{x})$ is zero at infinity) puts Eq.(\ref{gsolution}) into the form
 \begin{equation}
 \mathbf{g}(\mathbf{x})=G\int\frac{(\mathbf{x}^{\prime}-\mathbf{x})\rho_{c}(\mathbf{x}^{\prime})}{|\mathbf{x}-\mathbf{x}^{\prime}|^{3}}d^{3}x^{\prime},
 \label{Newg}
 \end{equation}
which is the same as the Newtonian result (\ref{gSolution}), but with only the inhomogeneous part $\rho_{c}(\mathbf{x})$ of the mass density contributing.  These considerations are summarized in:

\begin{quote}
\textbf{Result III:}  \textit{The Newtonian limit of the curvature equations is the vector Poisson equation}
\begin{equation}
\nabla^{2}\mathbf{g}=-4\pi G\nabla\rho_{g}
\label{VectorPoisson2}
\end{equation}
\textit{for the acceleration due to gravity $\mathbf{g}(\mathbf{x})$.  According to this equation, a homogeneous ``vacuum'' part $\rho_{qv}=constant$ of the source density $\rho_{g}(\mathbf{x})=\rho_{c}(\mathbf{x})+\rho_{qv}$ contributes not at all to the gravitational field $\mathbf{g}$, and, if the inhomogeneous part $\rho_{c}(\mathbf{x})$ of the source is nonzero only in some limited region of space, then this part of the source produces a gravitational field $\mathbf{g}(\mathbf{x})$ that is the same as if it were the only source in Newtonian gravitation theory, i.e., the curvature equations have the proper Newtonian limit.}
\end{quote}

\section{\label{sec:FirstInteg}The First Integral of the Curvature Equations}

In this section we consider to what extent solutions of the curvature equations differ from solutions of Einstein's equation. This difference becomes evident when we compare the first integral of the curvature equations to the conventional Einstein equation.  Both are Einstein equations but with different source terms.

We know from Result I that neither the energy-momentum tensor of the quantum vacuum $T^{\mu\nu}_{qv}$ nor the effective energy-momentum tensor $T^{\mu\nu}_{\Lambda}$ of the cosmological term contribute anything to solutions of the curvature equations, because these are nulled by the covariant derivatives in the source term (\ref{SourceTerm2}).  These tensors also play no role in the conservation law $\nabla_{\nu}T^{\mu\nu}=0$ of Condition 1, because again the covariant derivative nulls these parts of the full energy-momentum tensor $T^{\mu\nu}=T^{\mu\nu}_{c}+T^{\mu\nu}_{qv}+T^{\mu\nu}_{\Lambda}$.  For these reasons, $T^{\mu\nu}_{qv}$ and $T^{\mu\nu}_{\Lambda}$ simply do not appear in the curvature equations (\ref{CE2s}) or in the conservation law (\ref{Conservation}), and we may emphasize this fact by replacing $T^{\mu\nu}$ in these equations by $T^{\mu\nu}_{c}$.  But, as noted above, this raises the question of how a dark-energy contribution of the form $T^{\mu\nu}_{d}=-\rho_{d}c^{2}g^{\mu\nu}$ can cause accelerated expansion of the universe.

To begin to answer this question, we perform a contraction in the first curvature equation to obtain $\nabla_{\nu}R_{\alpha\beta}^{\ \ \  \mu\nu}=\nabla_{\beta}R_{\alpha}^{\ \mu}-\nabla_{\alpha}R_{\beta}^{\ \mu}$ (the contracted Bianchi identity), and then use this, together with the source term (\ref{SourceTerm2}), in the second curvature equation (\ref{SourceEq2}) to put the latter into the form
\begin{eqnarray}
& &\nabla_{\alpha}\left[R_{\beta}^{\ \mu}-\frac{8\pi G}{c^{4}}\left(T_{\beta}^{\ \mu}-\frac{1}{2}\delta^{\ \mu}_{\beta}T\right)\right] \nonumber  \\
& &-\nabla_{\beta}\left[R_{\alpha}^{\ \mu}-\frac{8\pi G}{c^{4}}\left(T_{\alpha}^{\ \mu}-\frac{1}{2}\delta^{\ \mu}_{\alpha}T\right)\right]=0.
\label{Ecurvature}
\end{eqnarray}
Incidentally, this form makes it obvious that, if $g_{\alpha\beta}$ is a solution of the conventional Einstein's equation, i.e., the quantities in square brackets vanish, it is also a solution of the curvature equations. 
Equation (\ref{Ecurvature}) shows again that the parts $T_{qv\ \mu}^{\ \ \ \ \nu}=-\rho_{qv}c^{2}\delta_{\mu}^{\ \nu}$ and $T_{\Lambda\ \mu}^{\ \ \ \nu}=-\rho_{\Lambda}c^{2}\delta_{\mu}^{\ \nu}$ of the energy-momentum tensor $T_{\mu}^{\ \nu}$ make no contribution to the curvature equations because the quantities in curved brackets for these parts are constant multiples of the metric tensor $\delta_{\mu}^{\ \nu}$, and these are nulled by the covariant derivatives in this equation.  Therefore $T_{\mu}^{\ \ \nu}=T_{c\ \mu}^{\ \ \ \nu}+T_{qv\ \mu}^{\ \ \ \nu}+T_{\Lambda\ \mu}^{\ \ \  \ \nu}$ may be replaced by $T_{c\ \mu}^{\ \ \ \nu}$ in Eq.~(\ref{Ecurvature}) to obtain
\begin{eqnarray}
& &\nabla_{\alpha}\left[R_{\beta}^{\ \mu}-\frac{8\pi G}{c^{4}}\left(T_{c\ \beta}^{\ \ \ \mu}-\frac{1}{2}\delta^{\ \mu}_{\beta}T_{c}\right)\right] \nonumber  \\
& &-\nabla_{\beta}\left[R_{\alpha}^{\ \mu}-\frac{8\pi G}{c^{4}}\left(T_{c\ \alpha}^{\ \ \ \mu}-\frac{1}{2}\delta^{\ \mu}_{\alpha}T_{c}\right)\right]=0.
\label{Ecurvature2}
\end{eqnarray}
This shows that a metric $g_{\alpha\beta}$ that satisfies an Einstein equation with only the classical energy-momentum tensor $T_{c}^{\mu\nu}$ acting as source term is also a solution of the curvature equations, but such is not the most general solution of the curvature equations.  If $g_{\alpha\beta}$ is a solution of the curvature equations that is \emph{not} a solution of $R_{\mu}^{\ \nu}=8\pi G(T_{c\ \mu}^{\ \ \ \nu}-\delta_{\mu}^{\ \nu}T_{c}/2)/c^{4}$, then the quantity
\begin{equation}
\frac{8\pi G}{c^{4}}Z_{\alpha}^{\ \mu}\equiv R_{\alpha}^{\ \mu}-\frac{8\pi G}{c^{4}}\left(T_{c\ \alpha}^{\ \ \ \mu}-\frac{1}{2}\delta_{\alpha}^{\ \mu}T_{c}\right)
\label{ZThing}
\end{equation}
does not vanish, and Eq.~(\ref{Ecurvature2}) may be written in terms of $Z_{\alpha}^{\ \mu}$ as
\begin{equation}
\nabla_{\alpha}Z_{\beta}^{\ \mu}-\nabla_{\beta}Z_{\alpha}^{\ \mu}=0.
\label{Zequation}
\end{equation}
Now $Z_{\alpha}^{\ \mu}$ can always be written as the Hilbert conjugate
\begin{equation}
Z_{\alpha}^{\ \mu}=X_{\alpha}^{\ \mu}-\frac{1}{2}\delta_{\alpha}^{\ \mu}X
\label{HXconjugate}
\end{equation}
of a tensor
\begin{equation}
X_{\alpha}^{\ \mu}=Z_{\alpha}^{\ \mu}-\frac{1}{2}\delta_{\alpha}^{\ \mu}Z,
\label{HZconjugate}
\end{equation}
where $Z=Z_{\mu}^{\ \mu}$ and $X=X_{\mu}^{\ \mu}$.  That is to say, the Hilbert conjugate contains the same information as the original tensor, and the Hilbert conjugate of a Hilbert conjugate is the original tensor.  Therefore, using (\ref{HXconjugate}) in Eq.~(\ref{ZThing}), noting that $R_{\alpha}^{\ \mu}=G_{\alpha}^{\ \mu}-\delta_{\alpha}^{\ \mu}G/2$ is the Hilbert conjugate of the Einstein tensor $G_{\alpha}^{\ \mu}$ (here $G=G_{\mu}^{\ \mu}$ is not the gravitational constant), and taking the Hilbert conjugate of the entire Eq.~(\ref{ZThing}) after these substitutions, we obtain
\begin{equation}
G_{\alpha}^{\ \mu}=\frac{8\pi G}{c^{4}}\left(T_{c\ \alpha}^{\ \ \ \mu}+X_{\alpha}^{\ \mu}\right),
\label{EinsteinLikeEq}
\end{equation}
and Eq.~(\ref{Zequation}), now an equation for $X_{\alpha}^{\ \mu}$, reads
\begin{equation}
\nabla_{\alpha}\left(X_{\beta}^{\ \mu}-\frac{1}{2}\delta_{\beta}^{\ \mu}X\right)=\nabla_{\beta}\left(X_{\alpha}^{\ \mu}-\frac{1}{2}\delta_{\alpha}^{\ \mu}X\right).
\label{Xcondition}
\end{equation}
We have shown that any solution of the curvature equations (\ref{CE2s})-(\ref{SourceTerm2}) is a solution of the Einstein equation (\ref{EinsteinLikeEq}) with an additive contribution $X_{\alpha}^{\ \mu}$ to the energy-momentum tensor $T_{c\ \alpha}^{\ \ \ \mu}$, and $X_{\alpha}^{\ \mu}$ must be a solution of Eq.~(\ref{Xcondition}).

The interpretation of Eq.~(\ref{EinsteinLikeEq}) is clear.  The curvature equations are differential equations containing derivatives of the metric one order higher than those in Eq.~(\ref{EinsteinLikeEq}).  Therefore (\ref{EinsteinLikeEq}) is the first integral of the curvature equations, and the $X_{\alpha}^{\ \mu}$ are somewhat arbitrary integration functions, subject  to condition (\ref{Xcondition}). We shall call $X_{\alpha}^{\ \mu}$ the  ``integration tensor'' for the curvature equations. 

 If we take the divergence of Eq.~(\ref{EinsteinLikeEq}), use the identity $\nabla_{\mu}G_{\alpha}^{\ \mu}=0$ and Condition 1, $\nabla_{\mu}T_{\alpha}^{\ \mu}=0$, we obtain a second condition on the integration tensor $X_{\alpha}^{\ \mu}$,
\begin{equation}
\nabla_{\mu}X_{\alpha}^{\ \mu}=0.
\label{Xcondition2}
\end{equation}
Then contracting on indices $\beta$ and $\mu$ in Eq.~(\ref{Xcondition}), and using (\ref{Xcondition2}) we have
\begin{equation}
\nabla_{\alpha}X=\partial_{\alpha}X=0,
\label{XderivZero}
\end{equation}
or
\begin{equation}
X=X_{\mu}^{\ \mu}=constant,
\label{Xconstant}
\end{equation}
and use of this in equation (\ref{Xcondition}) reduces the latter to the simpler form
\begin{equation}
\nabla_{\alpha}X_{\beta}^{\ \mu}=\nabla_{\beta}X_{\alpha}^{\ \mu}.
\label{Xcondition3}
\end{equation}

 Because the components $X_{\mu}^{\ \nu}$ are integration functions, they are determined by initial conditions or boundary conditions.  Specifically, if we know the curvature tensor $R^{\alpha}_{\ \beta\mu\nu}$ and the energy-monetum tensor $T_{c\ \mu}^{\ \ \ \nu}$ on a spacelike hypersurface $\Sigma$ at some initial coordinate time $t$, then the integration tensor is determined on $\Sigma$ by a rearrangement of Eq.~(\ref{EinsteinLikeEq}),
\begin{equation}
X_{\mu}^{\ \nu}=\frac{c{4}}{8\pi G}G_{\mu}^{\ \nu}-T_{c\ \mu}^{\ \ \ \nu},
\label{InitialCon}
\end{equation}
and the evolution of $X_{\mu}^{\ \nu}$ from these initial conditions is determined by Eqs.~(\ref{Xcondition2}) and (\ref{Xcondition3}), as shown in Appendix \ref{sec:XCauchy}.  These considerations lead us to the following result.

\begin{quote}
\textbf{Result IV:}  \textit{Solutions $g_{\mu\nu}$ of the curvature equations are solutions of the first integral of the curvature equations, or ``new Einstein equation'',}
\begin{equation}
G^{\mu\nu}=\frac{8\pi G}{c^{4}}\left(T^{\mu\nu}_{c}+X^{\mu\nu}\right) ;
\label{NewEE}
\end{equation}
\textit{in which the energy-momentum tensor $T^{\mu\nu}_{c}$ is augmented by an integration tensor $X^{\mu\nu}$.  The integration tensor $X_{\mu}^{\ \nu}$ is subject to conditions}
\begin{equation}
\nabla_{\nu}X_{\alpha}^{\  \nu}=0
\label{Xcondition22}
\end{equation}
\textit{and}
\begin{equation}
\nabla_{\alpha}X_{\beta}^{\ \mu}=\nabla_{\beta}X_{\alpha}^{\ \mu},
\label{Xcondition23}
\end{equation}
\textit{and is determined by the curvature tensor $R^{\alpha}_{\ \beta\mu\nu}$ and the energy-momentum tensor $T_{c\ \mu}^{\ \ \ \nu}$ at some initial time via a rearrangement of Eq.~(\ref{NewEE}),}
\begin{equation}
X_{\mu}^{\ \nu}=\frac{c{4}}{8\pi G}G_{\mu}^{\ \nu}-T_{c\ \mu}^{\ \ \ \nu},
\label{InitialCon2}
\end{equation}
\textit{and $X_{\mu}^{\ \nu}$  at later time is determined by the solution of Eqs.~(\ref{Xcondition22}) and (\ref{Xcondition23}) with these initial conditions.}
\end{quote}

\section{\label{sec:Solution}SOLUTIONS OF THE DARK-ENERGY AND VACUUM-ENERGY PROBLEMS}

We propose that the so-called ``dark energy'' is actually a manifestation of the integration tensor $X_{\mu}^{\ \nu}$.  It is a success of the curvature equations that an integration tensor of the form $X^{\mu\nu}=-\rho_{d}c^{2}g^{\mu\nu}$ is an allowed integration tensor [a solution of conditions (\ref{Xcondition22}) and (\ref{Xcondition23})], for this can answer the dark-energy puzzle.
The first integral of the curvature equations, namely the new Einstein equation (\ref{NewEE}), then takes the form
\begin{equation}
G^{\mu\nu}=\frac{8\pi G}{c^{4}}\left(T_{c}^{\mu\nu}-\rho_{d}c^{2} g^{\mu\nu}\right),
\label{FirstIntegral}
\end{equation}
which is known to agree with the observational data relating to the accelerated expansion of the universe if $\rho_{d}$ has the value $\rho_{d}=\Omega_{d}\rho_{cr}$ with $\Omega_{d}=0.7\pm 0.1$ [see Eq.~(\ref{DarkDensityEq})].

Because the components $X^{\mu\nu}$ are integration functions, $\rho_{d}c^{2}$ is an integration constant determined by initial conditions.   Specifically, the contraction of Eq.~(\ref{FirstIntegral})  gives the formula
\begin{equation} 
\rho_{d}c^{2}=\frac{1}{4}\left(T_{c}-\frac{c^{4}}{8\pi G}G_{\mu}^{\ \mu}\right),
\label{DarkDensity}
\end{equation}
for this constant, which can be evaluated at some initial point in time, or at \emph{any} time, since an integration constant is also a constant of the motion.

\begin{quote}
\textbf{Result V:} \textit{The first integral of the curvature equations corresponding to an integration tensor of the form $X^{\mu\nu}=-\rho_{d}c^{2}g^{\mu\nu}$, with $\rho_{d}=\Omega_{d}\rho_{cr}$ and $\Omega_{d}=0.7\pm 0.1$,}
\begin{equation}
G^{\mu\nu}=\frac{8\pi G}{c^{4}}\left(T_{c}^{\mu\nu}-\rho_{d}c^{2} g^{\mu\nu}\right),
\label{FirstIntegral2}
\end{equation}
\textit{accounts for the accelerated expansion of the universe.  The constant $\rho_{d}c^{2}$ is not the energy density of the quantum vacuum but an integration constant determined by initial conditions via  equation
\begin{equation}
\rho_{d}c^{2}=\frac{1}{4}\left(T_{c}-\frac{c^{4}}{8\pi G}G_{\mu}^{\ \mu}\right) ,
\label{InitialCondition}
\end{equation}
where $T_{c}=T_{c\ \mu}^{\ \ \ \mu}$.  The ``dark energy'' has the value it has because of the way our universe began, and could have a different value in a different universe.}
\end{quote}

The most general intergration tensor $X_{\mu}^{\ \nu}$ can be split into a traceless part $Y_{\mu}^{\ \nu}$ ($Y=Y_{\mu}^{\ \mu}=0$) and a part $-\rho_{d}c^{2}\delta_{\mu}^{\ \nu}$ determined by a single constant $\rho_{d}c^{2}=-X_{\mu}^{\ \mu}/4$ as
\begin{equation}
X_{\mu}^{\ \nu}=Y_{\mu}^{\ \nu}-\rho_{d}c^{2}\delta_{\mu}^{\ \nu}.
\label{DarkInitial}
\end{equation}
Because the energy-momentum of all material particles and field excitations is, by definition, contained in $T^{\mu\nu}_{c}$, the integration tensor $X^{\mu\nu}$ in the new Einstein equation (\ref{NewEE}) can only be interpreted as the Einstein vacuum $S^{\mu\nu}_{v}$.  Condition 2 then demands that this be of the Lorentz invariant form $X^{\mu\nu}=-\rho_{d}c^{2}g^{\mu\nu}$.  That is to say, Condition 2 demands that the traceless part of the integration tensor vanish, $Y_{\mu}^{\ \nu}=0$, in order that Eq.~(\ref{NewEE}) be consistent with the principle of equivalence.

Let us see specifically how Results IV and V solve the vacuum-energy problem, the dark-energy problem, and the cosmological-constant problem as defined in Section II.

\textbf{Vacuum-Energy Problem:}  The vacuum-energy problem exists because the conventional Einstein equation (\ref{StandardEquation}) contains the energy-momentum tensor of the quantum vacuum $T_{qv}^{\mu\nu}=-\rho_{qv}c^{2}g^{\mu\nu}$ as part of its source term.  As is well known, if reasonable estimates of the vacuum energy density $\rho_{qv}c^{2}$ are taken at face value and used as the source term in the conventional Einstein equation, we derive cosmological solutions that disagree with observation by many orders of magnitude.

The curvature equations, on the other hand, are immune from the vacuum energy problem because the vacuum energy of quantum fields has no effect on the curvature or metric of spacetime, according to these equations.  The first integral of the curvature equations, the new Einstein equation (\ref{NewEE}), does not contain the energy-momentum of the quantum vacuum  because this is a non-contributing energy-momentum tensor for the curvature equations\cite{Radiation}.

\textbf{Dark-Energy Problem:}   The ``dark-energy'' density $\rho_{d}c^{2}$ can be vastly smaller than the vacuum energy density of quantum fields because this quantity is an integration constant for the curvature equations determined by initial conditions, and is \emph{not} the vacuum energy density of any quantum field.  In fact, the term ``dark \emph{energy}'' is a misnomer.  Moreover, the small observed value of $\rho_{d}c^{2}$ is strong evidence that this is not the vacuum energy density of a quantum field.  The identification of the ``dark energy'' with an integration constant $\rho_{d}c^{2}$ determined by initial condition, carries with it the prediction that a quantum field with this energy density does not exist and will not be found in any search for such a field.

\textbf{Cosmological-Constant Problem:} The cosmological constant $\Lambda$ in Einstein's equation (\ref{EinsteinEq}) can be understood as a trivially different version of the integration constant $\rho_{d}c^{2}$ [$\Lambda=8\pi G(\rho_{d}c^{2})/c^{4}$], and as such it is determined by initial conditions for the curvature equations; it does not represent the vacuum energy density of any field, and the Einstein equation with cosmological term is to be understood as a first integral of the curvature equations. 

Our concentration on vacuum energy and integration tensors is still a step removed from the main point we should emphasize here.  We should keep in mind that the integration tensor $X^{\mu\nu}=-\rho_{d}c^{2}g^{\mu\nu}$ is merely the representation, in the first integral of the curvature equations, of certain initial conditions for the curvature equations.  The bare fact of the matter is this: \emph{The curvature equations admit solutions representing accelerated expansion even if the energy-momentum tensors $T^{\mu\nu}_{c}$, $T^{\mu\nu}_{qv}$, and $T^{\mu\nu}_{\Lambda}$ are all zero} (or perhaps with a negligible density of dust particles so that we can track the expansion). This hypothetical situation is not so different from the universe in which we live, for which the integration tensor $X^{\mu\nu}=-\rho_{d}c^{2}g^{\mu\nu}$ dominates the matter and radiation represented by $T_{c}^{\mu\nu}$ in the new Einstein equation (\ref{NewEE}). The Einstein equation of CGR, on the other hand, requires a vacuum energy or cosmological term to generate accelerated expansion.  This again emphasizes the fact that accelerated expansion is a matter of initial conditions for the curvature equations and \emph{not} a question of vacuum energy, and, when the curvature equations are adopted as the field equations of general relativity, no field of any kind is necessary to account for accelerated expansion, .

\section{\label{sec:Inflation}INFLATION}

Inflation is driven by the energy-momentum tensor
\begin{equation}
T^{\mu\nu}_{fv}=-\rho_{fv}c^{2}g_{\mu\nu} 
\label{FalseVacuum}
\end{equation}
of the ``false vacuum'' of grand unified theory \cite{Guth, Linde2, Blau, Linde3}.  Given that the form of this tensor is essentially the same as the vacuum energy-momentum tensor of quantum fields, one might expect that the ``false vacuum'', like the true vacuum, would contribute nothing to the source term $J_{\alpha\beta}^{\ \ \mu}$ of the curvature equations.  But this is \emph{not} the case.

The all important difference between the energy-momentum tensor (\ref{FalseVacuum}) of the false vacuum and the energy-momentum tensor $T^{\mu\nu}_{qv}=-\rho_{qv}c^{2}g^{\mu\nu}$ of a true vacuum is that $\rho_{fv}(t)$ is a function of time and $\rho_{qv}$ is a constant.  For the false vacuum, the source term of the curvature equations,
\begin{equation}
J_{\alpha\beta}^{\ \ \mu}=\frac{2G}{c^{3}}\frac{d\rho_{fv}}{dt}\left(\delta_{\alpha}^{\ 0}\delta_{\beta}^{\ \mu}-\delta_{\alpha}^{\ \mu}\delta_{\beta}^{\ 0}\right) ,
\label{FVSource}
\end{equation}
does not vanish, and this drives inflation.  The false-vacuum energy-momentum tensor is a contributing energy-momentum tensor and appears in the first integral of the curvature equations (\ref{FirstIntegral}) as part of $T^{\mu\nu}_{c}$ along with any other contributing energy-momentum tensors. Consequently inflation is driven by $T^{\mu\nu}_{fv}$ in essentially the same way as in CGR.  We learn from this example that whether an energy-momentum tensor is contributing or not depends not only on the form of the tensor at a given time but also on how the tensor evolves in time.

\section{\label{sec:Conclusion}CONCLUSION}

The present paper has as its goal to make plausible the idea that Nature has chosen the same basic pattern for her equations of gravitation and electrodynamics.  The central feature of this analogy is the set of curvature field equations, which are formally very similar to the Maxwell equations of electrodynamics.  A summary of results based on the curvature equations is unnecessary here because the important results are broken out in the text and are easily reviewed by scanning through the paper and reading only these.  It suffices here to emphasize the meaning of some of these results for general relativity.  What we hope the reader will carefully consider is whether the curvature equations are a more satisfactory choice for the field equations of general relativity than is the conventional Einstein equation.  The facts suggesting this might be the case are the following:

\subsection{The Case For the Curvature Equations}

\begin{itemize}
\item The curvature equations are immune from the vacuum-energy problem because the vacuum energy of quantum fields has no effect on the curvature or metric of spacetime, according to these equations.  The solution of the vacuum-energy problem is important, of course,  because, if the very large estimates of the vacuum energy density of quantum fields are correct and used as the source term in the conventional Einstein's equation, we derive what is perhaps the largest discrepancy between observation and an accepted theory yet encountered in physics.

\item  The curvature equations explain the accelerated expansion of the universe by identifying the  ``dark energy''  as a certain integration constant in the first integral of the curvature equations, which is unrelated to the vacuum energy of quantum fields.  This allows us to understand how the ``dark energy density'' can be so vastly smaller than estimates of the vacuum energy density of quantum fields, and shows the term ``dark \emph{energy}'' to be a misnomer.  Moreover, the observed small value of the ``dark-energy density'' is strong evidence that the ``dark-energy'' is not the vacuum energy of a quantum field.

\item With the ``dark energy'' identified as an integration tensor, no field is necessary to drive the accelerated expansion of the universe.  An accelerating universe is simply one of the solutions of the curvature equations, even for an empty universe containing no particles or radiation whatsoever.  Therefore, the interpretation of general relativity proposed here predicts that no field will be found whose vacuum energy density has the observed value of $\rho_{d}c^{2}$ because none is necessary.

\item  The curvature equations also predict that the effective cosmological constant $\Lambda=8\pi G\rho_{d}/c^{2}$ is \emph{exactly constant}.  An observation showing a time-varying cosmological constant, as would be expected for a ``slow-roll'' scalar field, would rule out the present interpretation of dark energy.

\item  The curvature equations, together with Conditions 1 and 2, are equivalent to the new Einstein equation (\ref{FirstIntegral2}), and this is known to be consistent with all observational tests of general relativity.  The new Einstein equation, with no contribution from the quantum vacuum energy and with the dark-energy density (or cosmological constant) chosen to agree with observation, is, in fact, the working field equation of present-day general relativity. 
\end{itemize}

\subsection{The Case Against the Curvature Equations}

In addition to the above arguments suggesting that the curvature equations might better serve as the field equations of general relativity than the original Einstein equation, there are, of course, arguments against adopting the curvature equations for this purpose.  We list below some of these objections:
\begin{itemize}
\item  Without additional conditions, the curvature equations have vastly more solutions than the conventional Einstein equation.  Many of these solutions (which are associated with the different possible integration tensors $X^{\mu\nu}$ in Result IV) are unphysical or are inconsistent with observation.  

\item  The \emph{local} conservation of energy-momentum, as expressed by the condition $\nabla_{\nu}T^{\mu\nu}=0$, is not an immediate consequence of the curvature equations (as it is for the conventional Einstein field equation), and must be imposed as a additional condition (Condition 1), albeit a condition required by the principle of equivalence.

\item  The vacuum source term $X^{\mu\nu}$ in the new Einstein equation (\ref{NewEE}), which is derived from the curvature equations, is not automatically of the locally Lorentz invariant form $X^{\mu\nu}=-\rho_{d}c^{2}g^{\mu\nu}$.  This too must be imposed as an additional constraint (Condition 2) if solutions of the curvature equations are to be consistent with the principle of equivalence and with observation.

\item  Even with Conditions 1 and 2 accepted as reasonable, the value of the dark-energy density $\rho_{d}c^{2}$ is not predicted by the curvature equations; it is an integration constant determined by the initial conditions for the universe; conditions which could be different in a universe that began differently.  Therefore we have no predicted value for the dark-energy density to be compared with the observed value as evidence for or against the curvature equations.
\end{itemize}

In summary, on the one hand the curvature equations have a \emph{particular} first integral,
\begin{equation}
G^{\mu\nu}=\frac{8\pi G}{c^{4}}\left(T_{c}^{\mu\nu}-\rho_{d}c^{2} g^{\mu\nu}\right),
\label{FirstIntegral22}
\end{equation}
which is an Einstein equation consistent with \emph{all} observational reults.  The vacuum energy of quantum fields does not appear in this equation because the vacuum energy-momentum tensor contributes nothing to the source term of the curvature equations, and the ``dark energy density'' $\rho_{d}c^{2}$ can have its small observed value because this is an integration constant for the curvature equations unrelated to the large vacuum energy density.  This could possibly be the answer to the vacuum-energy and dark-energy puzzles of general relativity [and Eq.~(\ref{FirstIntegral22}) is a first integral of the curvature equations, and a possible solution to these problems, whether or not we impose Conditions 1 and 2 ].

On the other hand, Eq.~(\ref{FirstIntegral22}) is not a \emph{unique} first integral of the curvature equations.  There are many other possible first integrals, and among these many that disagree with observation.  We have argued that Conditions 1 and 2 (which we view as consequences of the principle of equivalence) go a long way toward removing this arbitrariness and restricting  the general first integral of Result IV to the form (\ref{FirstIntegral22}).  But the introduction of such subsidary conditions does not seem entirely natural or satisfying.  In the final analysis, we cannot say why Nature has chosen the particular first integral of the curvature equations (\ref{FirstIntegral22})  for the universe in which we live, only that it is a possibility she has done so.

\begin{acknowledgments}
I  am indebted to my teacher, Professor Edward Teller (deceased), who attempted to teach me how to think imaginatively about physics.  Credit also goes to Hans C. Ohanian who provided helpful comments on a very early version of the manuscript, and to M. Shane Burns whose calculations ruled out a certain speculative solution of the curvature equations.  I must also thank Heidi Fearn, Peter W. Milonni, Rita Cook, Donald R. Erbschloe, James H. Head, Randell J. Knize, Iyad A. Dajani, and Brian M. Patterson for helpful comments on this work.
\end{acknowledgments}

\appendix

\section{\label{sec:CEderivation} Derivation of the Curvature Equations from the Einstein Equation}

The first of the curvature equations (\ref{Bianchi}) is the Bianchi 
relation.  It is a well known identity derivable from the definition 
of the curvature tensor and does not require proof here.

The second curvature equation (\ref{SourceEq}) is derived from the Einstein equation as follows.  Raise the index $\alpha$ in (\ref{Bianchi}), 
contract this index with $\omega$, and use the definition $R_{\beta\mu}= 
{R^{\omega}}_{\beta\omega\mu}$ of the Ricci tensor to obtain the contracted Bianchi identity
\begin{equation}
\nabla_{\omega}{R^{\omega}}_{\beta\mu\nu}=\nabla_{\mu}R_{\beta\nu}
-\nabla_{\nu}R_{\beta\mu}.
\label{Identity}
\end{equation}
Then use the Einstein equation
\begin{equation}
R_{\beta\nu}=\frac{8\pi G}{c^{4}}(T_{\beta\nu}-\frac{1}{2}g_{\beta\nu}T)
\label{EEver2}
\end{equation}
on the right in (\ref{Identity}) and the symmetry properties of the 
curvature tensor 
(${R^{\omega}}_{\beta\mu\nu}=-{R_{\mu\nu\beta}}^{\omega}$) to obtain
\begin{eqnarray}
\nabla_{\omega}{R_{\mu\nu\beta}}^{\omega}&=&-\frac{8\pi G}{c^{4}}[
\nabla_{\mu}(T_{\nu\beta}-\frac{1}{2}g_{\nu\beta}T) \nonumber \\
& & \mbox{} \ \ \ \ \ \ -\nabla_{\nu}(T_{\mu\beta}-\frac{1}{2}g_{\mu\beta}T)].
\label{CurveEq}
\end{eqnarray}
Raising of index $\beta$ and relabeling indices puts this in the 
form of the second curvature equation (\ref{SourceEq}) with source term 
(\ref{SourceTerm}).  

Incidentally, if we include a cosmological term $\Lambda 
g_{\beta\nu}$ on the right 
in the Einstein equation (\ref{EEver2}), the result (\ref{CurveEq}) is 
unchanged because the covariant derivatives on the right in 
(\ref{Identity}) gives zero when acting on the cosmological term.

\section{\label{sec:Action} Action Principles: Gravitational and Electrodynamic}

The action for the curvature equations with a prescribed source  $J_{\mu}^{\ \alpha\beta}$,
\begin{eqnarray}
S_{G}&=&\frac{-1}{16\pi}\int_{\Sigma}(R^{\alpha\beta\mu\nu}R_{\alpha\beta\mu\nu} \nonumber \\
& &\mbox{}+16\pi J_{\mu}^{\ \alpha\beta}\Gamma^{\mu}_{\ \alpha\beta})\sqrt{-g}\ d^{4}x,
\label{GravAction2}
\end{eqnarray}
is expressed in terms of the ``metric free'' form of the curvature tensor,
\begin{equation}
{R^{\alpha}}_{\beta \mu \nu} =
\partial_{\mu}\Gamma^{\alpha}_{\ \beta \nu}-
\partial_{\nu}\Gamma^{\alpha}_{\ \beta \mu}+
\Gamma^{\alpha}_{\ \tau \mu}\Gamma^{\tau}_{\ \beta \nu}-
\Gamma^{\alpha}_{\ \tau \nu}\Gamma^{\tau}_{\ \beta \mu},
\label{CurveTens2}
\end{equation}
as
\begin{eqnarray}
S_{G}&=&\frac{-1}{16\pi}\int_{\Sigma}(g^{\beta\gamma}g^{\mu\tau}g^{\nu\delta}g_{\alpha\omega}R^{\alpha}_{\ \gamma\tau\delta}R^{\omega}_{\ \beta\mu\nu} \nonumber \\
& &\mbox{}+16\pi J_{\mu}^{\ \alpha\beta}\Gamma^{\mu}_{\ \alpha\beta})\sqrt{-g}\ d^{4}x.
\label{NewAction}
\end{eqnarray}

This is analogous to expressing the electromagnetic action,
\begin{eqnarray}
S_{EM}&=&\frac{-1}{16\pi}\int_{\Sigma}(F^{\mu\nu}F_{\mu\nu}\nonumber \\
& &\mbox{}+16\pi J^{\mu}A_{\mu})\sqrt{-g}\ d^{4}x,
\label{EMAction}
\end{eqnarray}
in terms of the ``metric free''  form of the field tensor,
\begin{equation}
F_{\mu\nu}=\nabla_{\mu}A_{\nu}-\nabla_{\nu}A_{\mu},
\label{Field}
\end{equation}
as
\begin{eqnarray}
S_{EM}&=&\frac{-1}{16\pi}\int_{\Sigma}(g^{\alpha\mu}g^{\beta\nu}F_{\alpha\beta}F_{\mu\nu}\nonumber \\& &\mbox{}+16\pi J^{\mu}A_{\mu})\sqrt{-g}\ d^{4}x
\label{EMAction2}
\end{eqnarray}
in the derivation of Maxwell's equations.

In the electromagnetic derivation, one then considers a variation $\delta A_{\mu}$ of the potentials $A_{\mu}$ that is arbitrary except that it vanishes on the boundary of the four-volume $\Sigma$.  The variation of (\ref{EMAction2}) is
\begin{eqnarray}
\delta S_{EM}&=&\frac{-1}{16\pi}\int_{\Sigma}(g^{\alpha\mu}g^{\beta\nu}[(\delta F_{\alpha\beta})F_{\mu\nu}+F_{\alpha\beta}(\delta F_{\mu\nu})]\nonumber \\
& &\mbox{}+16\pi J^{\mu}\delta A_{\mu})\sqrt{-g}\ d^{4}x,
\label{VaryEMAction2}
\end{eqnarray}
where
\begin{equation}
\delta F_{\mu\nu}=\nabla_{\mu}\left(\delta A_{\nu}\right)-\nabla_{\nu}\left(\delta A_{\mu}\right).
\label{FieldVariation}
\end{equation}

Essentially the same procedure is used in the gravitational case.  The change of (\ref{NewAction}), for a variation $\delta\Gamma^{\mu}_{\ \alpha\beta}$ of the connections, treating $J_{\mu}^{\ \alpha\beta}$ as a prescribed source term,  reads
\begin{eqnarray}
\delta S_{G}&=&\frac{-1}{16\pi}\int_{\Sigma}(g^{\beta\gamma}g^{\mu\tau}g^{\nu\delta}g_{\alpha\omega}[(\delta R^{\alpha}_{\ \gamma\tau\delta})R^{\omega}_{\ \beta\mu\nu}\nonumber \\
& &\mbox{}+ R^{\alpha}_{\ \gamma\tau\delta}(\delta R^{\omega}_{\ \beta\mu\nu})]\nonumber \\
& &\mbox{}+16\pi J_{\mu}^{\ \alpha\beta}\delta\Gamma^{\mu}_{\ \alpha\beta})\sqrt{-g}\ d^{4}x,
\label{VaryNewAction}
\end{eqnarray}
where the variation of the curvature tensor is given by the Palatini identity,
\begin{equation}
\delta R^{\alpha}_{\ \beta\mu\nu}=\nabla_{\mu}\left(\delta \Gamma^{\alpha}_{\ \beta\nu}\right)-\nabla_{\nu}\left(\delta \Gamma^{\alpha}_{\ \beta\mu}\right),
\label{Palitine2}
\end{equation}
which is the gravitational analogue of Eq.~(\ref{FieldVariation}).

In both cases the ``metric free'' form of the field tensor represents the solution of the source-free field equation.  The expression (\ref{Field}) for the field tensor $F_{\mu\nu}$ in terms of the potentials $A_{\mu}$ is the solution of the Maxwell equation 
\begin{equation}
\nabla_{\omega}F_{\mu \nu}+\nabla_{\nu}F_{\omega 
\mu}+\nabla_{\mu}F_{\nu \omega}= 0,
\label{Max1Two}
\end{equation}
and the formula (\ref{CurveTens2}) for the gravitational field $R^{\alpha}_{\ \beta\mu\nu}$ in terms of the ``potentials'' $\Gamma^{\mu}_{\ \alpha\beta}$ is the solution of  the first curvature equation,
\begin{equation}
 \nabla_{\omega}R^{\alpha}_{\  \beta \mu \nu}+
 \nabla_{\nu}R^{\alpha}_{\  \beta \omega \mu}+
  \nabla_{\mu}R^{\alpha}_{\ \beta \nu \omega}=0 .  
  \label{Bianchi22}
\end{equation}

Continuing in this parallel fashion, we use (\ref{FieldVariation}) in Eq.~(\ref{VaryEMAction2}), invoke the antisymmetry of $F_{\mu\nu}$, and perform some index manipulations to obtain
\begin{eqnarray}
\delta S_{EM}&=&\frac{-1}{4\pi}\int_{\Sigma}[-F^{\mu\nu}\nabla_{\nu}(\delta A_{\mu})\nonumber \\
& &\mbox{}+4\pi J^{\mu}\delta A_{\mu}]\sqrt{-g}\ d^{4}x,
\label{Vary2}
\end{eqnarray}
and similarly for gravitation, using (\ref{Palitine2}) in Eq.(\ref{VaryNewAction}), and then using the symmetry of $R^{\alpha}_{\ \beta\mu\nu}$ on its last two indexes, we arrive at
\begin{eqnarray}
\delta S_{G}&=&\frac{1}{4\pi}\int_{\Sigma}[-R_{\alpha}^{\ \beta\mu\nu}\nabla_{\nu}(\delta\Gamma^{\alpha}_{\ \beta\mu})\nonumber \\
& &\mbox{}+4\pi J_{\mu}^{\ \alpha\beta}\delta\Gamma^{\mu}_{\ \alpha\beta}]\sqrt{-g}\ d^{4}x.
\label{GravVary}
\end{eqnarray}
Then using identities
\begin{eqnarray}
F^{\mu\nu}\nabla_{\nu}\left(\delta A_{\mu}\right)&=&\nabla_{\nu}\left[ F^{\mu\nu}\delta A_{\mu}\right]\nonumber \\
& &\mbox{}-\left(\nabla_{\nu}F^{\mu\nu}\right)\delta A_{\mu}, \label{EMident}  \\
R_{\alpha}^{\ \beta\mu\nu}\nabla_{\nu}\left(\delta\Gamma^{\alpha}_{\ \beta\mu}\right)&=&\nabla_{\nu}\left[R_{\alpha}^{\ \beta\mu\nu}\delta\Gamma^{\alpha}_{\ \beta\mu}\right]\nonumber \\
& &\mbox{}-\left(\nabla_{\nu}R_{\alpha}^{\ \beta\mu\nu}\right)\delta\Gamma^{\alpha}_{\ \beta\mu},
\label{Identities}
\end{eqnarray}
in Eqs.~(\ref{Vary2}) and (\ref{GravVary}), respectively, and applying Stokes' theorem
\begin{equation}
\int_{\Sigma}\nabla_{\nu}V^{\nu}\sqrt{-g}\ d^{4}x=\int_{\partial\Sigma}n_{\nu}V^{\nu}\sqrt{\gamma}\ d^{3}y,
\label{Stokes}
\end{equation}
which is applicable because the quantities in square brackets in (\ref{EMident}) and (\ref{Identities}) are vectors [here $\gamma_{ij}$ is the metric and $y^{i}$ the coordinates on the boundary $\partial\Sigma$ of $\Sigma$], we have vanishing boundary integrals because $\delta A_{\mu}$ and $\delta\Gamma^{\mu}_{\ \alpha\beta}$ vanish on the boundary, and (\ref{Vary2}) and (\ref{GravVary}) become
\begin{equation} 
\delta S_{EM}=\frac{-1}{4\pi}\int_{\Sigma}\left(\nabla_{\nu}F^{\mu\nu}+4\pi J^{\mu}\right)\delta A_{\mu}\sqrt{-g}d^{4}x,
\label{FinalEM}
\end{equation}
and
\begin{equation}
\delta S_{G}=\frac{-1}{4\pi}\int_{\Sigma}\left(\nabla_{\nu}R_{\alpha}^{\ \beta\mu\nu}+4\pi J_{\alpha}^{\ \beta\mu}\right)\delta\Gamma^{\alpha}_{\beta\mu}\sqrt{-g}d^{4}x.
\label{GravFinal}
\end{equation}
Finally, because $\delta A_{\mu}$ and $\delta\Gamma^{\alpha}_{\ \beta\mu}$ are arbitrary, the action principles $\delta S_{EM}=0$ and $\delta S_{G}=0$ give the second Maxwell equation
\begin{equation} \nabla_{\nu}F^{\mu\nu}=-4\pi J^{\mu}
\label{FinalMaxwell}
\end{equation}
and the second curvature equation
\begin{equation}
\nabla_{\nu}R_{\alpha}^{\ \beta\mu\nu}=-4\pi J_{\alpha}^{\ \beta\mu}.
\label{FinalGrav}
\end{equation}
for the prescribed sources under consideration. 

This derivation of the curvature equations from an action principle is not entirely satisfactory.  It is a valid derivation when $J_{\alpha\beta}^{\ \ \mu}$ is treated as a prescribed source term, i.e., $J_{\alpha\beta}^{\ \ \mu}$ is assumed not to change under the variation $\delta\Gamma^{\mu}_{\ \alpha\beta}$.  But, if the energy-momentum tensor $T^{\mu\nu}$ is taken to be the prescribed source instead of  $J_{\alpha\beta}^{\ \ \mu}$, then, according to its definition (\ref{SourceTerm2}), $J_{\alpha\beta}^{\ \ \mu}$ contains the connection coefficients $\Gamma^{\mu}_{\ \alpha\beta}$ because of the covariant derivatives in this definition, and so $J_{\alpha\beta}^{\ \ \mu}$ is not independent of the variation $\delta J_{\alpha\beta}^{\ \ \mu}$ and one does not obtain the correct source term in the curvature equations by this variation.  This difficulty is limited to the source term in the action (\ref{GravAction2}), and there is no problem with the action for the vacuum curvature equations.  In any case, this is a problem with the derivation of the curvature equations from an action principle, and not a problem with the curvature equations themselves, which are certainly correct equations in general relativity.

\section{\label{sec:XCauchy} The Cauchy Problem for the Integration Tensor}

For the present argument, we can work in Gaussian normal coordinates, for which the metric takes the form
\begin{equation}
ds^{2}=-(dx^{0})^{2}+g_{ij}dx^{i}dx^{j},
\label{GaussianNormal}
\end{equation}
 $x^{0}=ct$ is the time coordinate, and $x^{i}\ (i=1,\ 2,\ 3)$ are space coordinates.  For these ``synchronous coordinates'', condition (\ref{Xcondition22}) for the integration tensor $X^{\mu\nu}$ may be expanded as
\begin{equation}
\frac{1}{c}\frac{\partial X_{\mu}^{\ 0}}{\partial t}=-\frac{\partial X_{\mu}^{\ i}}{\partial x^{i}}+\Gamma^{\sigma}_{\ \nu\mu}X_{\sigma}^{\ \nu}-\Gamma^{\nu}_{\ \nu\sigma}X_{\mu}^{\ \sigma},
\label{Xcon1Expand}
\end{equation}
and, for $\alpha=0$ and $\beta=i$, condition (\ref{Xcondition23}) may be written as
\begin{equation}
\frac{1}{c}\frac{\partial X_{i}^{\ \mu}}{\partial t}=\frac{\partial X_{0}^{\ \mu}}{\partial x^{i}}-\Gamma^{\mu}_{\ 0\sigma}X_{i}^{\ \sigma}+\Gamma^{\mu}_{\ i\sigma}X_{0}^{\ \sigma}.
\label{Xcon2Expand}
\end{equation}
Thus, if $X^{\mu\nu}$ and its spatial derivatives are known on the spacelike hypersurface $\Sigma$ at a given value of time $t$, Eq.~(\ref{Xcon1Expand}) and (\ref{Xcon2Expand}) determine $X_{\mu}^{\ 0}(t+dt) [=X_{\mu}^{\ 0}(t)+(\partial X_{\mu}^{\ 0}/\partial t)dt]$ and $X_{i}^{\ \mu}(t+dt)$ [$=X_{i}^{\ \mu}(t)+(\partial X_{i}^{\ \mu}/\partial t)dt$] at a short time $dt$ into the future.  The remaining elements $X_{0}^{\ i}$ of $X_{\mu}^{\ \nu}$ are determined at time $t+dt$ by $X_{0}^{\ i}=-g^{ik}X_{k}^{\ 0}$, because $X_{\mu\nu}$ is symmetric, and we can step out the solution in this way for any number of time steps $dt$.
This demonstrates that, if $X_{\mu}^{\ \nu}$ and its spatial derivatives $\partial X_{\mu}^{\ \nu}/\partial x^{i}$ are known on $\Sigma$ at time $t_{0}$, the conditions (\ref{Xcondition22}) and (\ref{Xcondition23}) determine $X_{\mu}^{\ \nu}$ at later time $t$ in these coordinates, and in any coordinates by the tensor transformation law.


\begin{thebibliography}{99}
\bibitem{NormalOrder} It is common practice in quantum field theory to normally order the creation and annihilation operators making up the observable energy-momentum tensor.  This process is equivalent to discarding a formally infinite vacuum energy and has the effect of  assigning the value zero to the vacuum energy.  The reference point for measuring energy is indeed arbitrary in special-relativistic quantum field theory, but not in gravitation theory, where energy is the source of gravitational field.  The argument is made in quantum field theory that, in quantizing a classical field, the ordering of c-number variables that will become operators upon quantization is arbitrary to begin with, and therefore we can start with a classical ordering of variables in the energy-momentum tensor that upon quantization (upon specifying commutation rules) will give a normally ordered result, in which case the large vacuum energy never appears as part of the quantum energy-momentum tensor.  Is this a solution to the vacuum-energy problem?  The answer has to be ``no'', because different operator orderings give difference values for the vacuum energy, and, so long as the reference level for energy measurement is arbitrary in quantum field theories, such theories place no constraint whatever on the value of the vacuum energy.  One must invoke other arguments, such as giving credence to the reality of the zero-point energy of field oscillators, which would seem to be necessary to explain the Casimir effect, in order to estimate the vacuum energy density, and, generally speaking, such estimates turn out to be very large.
\end{thebibliography}
\end{document}